\documentclass[prd,twocolumn,a4paper,superscriptaddress,floatfix]{revtex4}
\usepackage{graphicx}
\begin{document}
%My commands
\newcommand{\be}{\begin{equation}}
\newcommand{\ee}{\end{equation}}
\newcommand{\bq}{\begin{eqnarray}}
\newcommand{\eq}{\end{eqnarray}}
\newcommand{\bsq}{\begin{subequations}}
\newcommand{\esq}{\end{subequations}}
\newcommand{\bc}{\begin{center}}
\newcommand{\ec}{\end{center}}
\newcommand\lapp{\mathrel{\rlap{\lower4pt\hbox{\hskip1pt$\sim$}} \raise1pt\hbox{$<$}}}
\newcommand\gapp{\mathrel{\rlap{\lower4pt\hbox{\hskip1pt$\sim$}} \raise1pt\hbox{$>$}}}

\def\E{{\cal E}}

\title{Evolution of Local and Global Monopole Networks}
\author{C.J.A.P. Martins}
\email[Electronic address: ]{Carlos.Martins@astro.up.pt}
\affiliation{Centro de Astrof\'{\i}sica, Universidade do Porto, Rua das Estrelas, 4150-762 Porto, Portugal}
\affiliation{DAMTP, University of Cambridge, Wilberforce Road, Cambridge CB3 0WA, United Kingdom}
\author{A. Ach\'ucarro} 
\email[Electronic address: ]{achucar@lorentz.leidenuniv.nl}
\affiliation{Institute Lorentz of Theoretical Physics, University of Leiden, 2333CA Leiden, The Netherlands}
\affiliation{Department of Theoretical Physics, University of the Basque Country UPV-EHU, 48940 Leioa, Spain} 
\date{16 June 2008}
\begin{abstract}
  We present an extension of the velocity-dependent one-scale model
  for cosmic string evolution, which is suitable for describing the
  evolution of local and global monopole networks. We
  discuss the key dynamical features that need to be accounted for, in
  particular the fact that the driving force is due to the other
  monopoles (rather than being due to local curvature as in the case
  of extended objects) and new forms of energy loss terms due to
  monopole-antimonopole capture and annihilation. For the case of
  local monopoles we recover and generalize the results of Preskill,
  suggesting that the scaling law for the monopole correlation length
  is very sensitive to the annihilation rate. On the other hand, for
  global monopoles the long-range forces generically lead
  to linear scaling (just like in the case of local cosmic
  strings). In this case we also find good qualitative agreement
  between our results and the numerical simulations of Bennett \& Rhie
  and Yamaguchi, although future high-resolution simulations will be
  needed for quantitative comparisons.
\end{abstract}
\pacs{98.80.Cq, 11.27.+d, 98.80.Es}
\maketitle

%%%%%%%%%%%%%%%%%%%%%%%%%%%%%%%%%%%%%%%%%%%%%%%%%%%%%%%%%%%%%%%%%%%%%%%%%%
\section{\label{intro}Introduction}

Since the pioneering work of Kibble \cite{KIBBLE} it has been known
that topological defects necessarily form at cosmological phase
transitions if they are stable. The details of the phase transition, in particular the
specific symmetry being broken, will determine what kind of defect
network is formed, and whether or not the network is long-lived.
Understanding defect formation, dynamical properties and evolution is
therefore a key part of any effort to understand the early universe. A
thorough overview of the subject can be found in the book by Vilenkin
and Shellard \cite{VSH}.

In the past three decades, most of the work on defects has focused on
cosmic strings, for the good reason that they are usually
cosmologically benign and are a generic prediction of inflationary
models based on Grand Unified Theories
\cite{Jeannerot:1997is,Jeannerot:2003qv} or branes
  \cite{Sarangi:2002yt, Majumdar:2002hy}. On the other hand, domain
  walls and monopoles are cosmologically quite dangerous, so any
  models in which they arise are subject to very tight bounds.
  Nevertheless, understanding their detailed properties is still
  important. On the one hand, in order to confidently impose bounds on
  models which form them one must know in detail how they evolve. On
  the other hand it is becoming increasingly clear, particularly in
  the context of models with extra dimensions such as brane inflation,
  that hybrid defect networks will often be produced. Two examples are
  semilocal strings and cosmic necklaces \cite{Dasgupta:2004dw,
  Chen:2005ae, Dasgupta:2007ds}. 
  To understand the evolution of these hybrid systems
  one needs to understand not only the dynamics of the strings but
  also that of the monopoles. This article is a first contribution to
  this task.

  The existence of magnetic monopoles was first suggested by Dirac in
  1931 \cite{DIRAC}, but first demonstrated in a field theory context
  only in 1974 by 't Hooft and Polyakov \cite{THOOFT,POLYAK}. It soon
  became apparent that they are cosmologically disastrous, and indeed
  this has been given the status of a 'problem'---the monopole problem
  \cite{VSH,KOLB}. For this reason, their study has been comparatively
  neglected relative to that of cosmic strings and even domain walls.

  Global monopoles have logarithmically divergent energy due to the
  slow fall-off of angular gradients, which makes some of their
  properties counterintuitive. Even their stability was the subject of
  some debate because a global monopole with its core position fixed
  has an angular zero mode \cite{Goldhaber} and a finite energy
  barrier between topologically distinct sectors \cite{AchUrr}. In
  reality, any angular deformation away from spherical symmetry leads
  to a displacement of the core and the only decay channel left is
  monopole annihilation \cite{Rhie,Perivolaropoulos}. Global monopoles
  are known to have scaling solutions \cite{BRHIE,YAMAGUCHI},
and could play a role in structure formation that can be constrained
with CMB data \cite{cmb1,cmb2} (see also \cite{cmb3}).
 
  In this paper we propose an extension of the velocity-dependent
  one-scale model for string evolution \cite{MS1,MS2} that can be used
  to study the evolution of local or global monopole networks. In
  future work we shall discuss how this can be applied to the study of
  hybrid defect networks.

  The rest of the article is organized as follows. We start in Sect.
  \ref{modelgener} by presenting a simplified derivation of the
  evolution equations for the two dynamical quantities of our model, a
  length scale that can be thought of as a correlation length and a
  root-mean squared (RMS) velocity. We then focus on the specific case
  of monopoles, first by providing (in Sect. \ref{monop}) a brief
  overview of previous work on monopole properties and dynamics and
  then by discussing (in Sect. \ref{modelspec}) how these features can
  be encapsulated in our model. In Sect. \ref{local} we use the model
  to discuss the evolution of networks of local monopoles; we shall
  see that our results agree with but generalize those of Preskill
  \cite{PRESKILL}. Similarly, the evolution of global monopole
  networks is discussed in Sect. \ref{global}; here we find good
  qualitative agreement with previous numerical simulations by Bennett
  and Rhie \cite{BRHIE} and by Yamaguchi \cite{YAMAGUCHI}, although
  the relatively low resolution of these simulations does not allow
  quantitative comparisons (which must be left for future work).
  In Sect. \ref{whatif} we revisit
some our modeling assumptions on the energy loss terms and driving forces,
and discuss how these affect our results.
  Finally in Sect. \ref{concl} we summarize our results and discuss
  some open issues.

%%%%%%%%%%%%%%%%%%%%%%%%%%%%%%%%%%%%%%%%%%%%%%%%%%%%%%%%%%%%%%%%%%%%%%%%%%
\section{\label{modelgener}Analytic Modeling Generalities}

Broadly speaking, the key idea behind the analytic modeling of defect
networks is that one give up the idea of studying their `statistical
physics' (which to a large extent can only be done numerically) and
instead concentrates on its `thermodynamics'. In other works, one or
more macroscopic quantities are chosen to describe the network, and
the knowledge of the microphysics is used to derive evolution equation
for these macroscopic quantities. This has the advantage of leading to
relatively simple models which can in principle encapsulate most of
the relevant physics, but it also has an associated cost: in going
from the microphysics to the macrophysics one is forced to introduce
phenomenological parameters, which can only be calibrated by direct
comparison with numerical simulations.

The first effort along these lines is Kibble's one-scale model of
string networks \cite{KIB} (extended in \cite{BMOD}), which has a
single macroscopic parameter: a lengthscale that can be identified as
the string correlation length, the string curvature radius, or the
inter-string distance---in the model's approximation they should be
seen as identical or at least comparable. This was later generalized
to a three-scale model by Austin, Copeland and Kibble \cite{ACK} where
there are three distinct lengthscales: the first two are again the
string correlation length and the inter-string distance, while the
third is a typical scale for small-scale `wiggles'.

A different approach recognizes that in order to be able to
quantitatively describe the whole cosmological history of these
networks one must be able to describe the evolution of the defect
velocities, not least because depending on the cosmological epoch and
on their own properties, the defects may be moving at non-relativistic
or at ultra-relativistic speeds. This is the basis for the
velocity-dependent one-scale model of Martins and Shellard
\cite{MS1,MS2}, which retains Kibble's assumptions on the existence of
a single lengthscale but adds the RMS velocity as a second macroscopic
quantity.  This model can be accurately derived starting from the
Goto-Nambu action, and it has also been successfully tested against
high-resolution field theory \cite{ABELIAN} and Goto-Nambu numerical
simulations \cite{MS3,MS4}. More recently, it has also been extended
to domain walls \cite{AWALL}. In what follows we provide a simplified
derivation of the dynamical equations of the model, along the lines of
\cite{AWALL}.

Consider a network of defects with $n$-dimensional worldsheets ($n=1$
for monopoles) evolving in $(3+1)$ space-time dimensions.  Temporarily
assume them to have velocity $v$, to be non-interacting and (for
extended objects) planar. Then the momentum per unit comoving defect
volume --simply the momentum, for monopoles-- goes as \be p\propto
a^{-1}\Longrightarrow v\gamma\propto a^{-n} \ee from which we get by
differentiation \be \frac{dv}{dt}+nH(1-v^2)v=0\,.  \ee On the other
hand, under the above hypotheses the average number of defects in a
fixed comoving volume should be conserved, which implies \be
\rho\propto \gamma a^{-(4-n)} \ee and again, differentiating and using
the velocity equation, we get \be
\frac{d\rho}{dt}+H[(4-n)+nv^2]\rho=0\,.  \ee The hypotheses so far
are, of course, widely unrealistic. However, we can use this as a
starting point to build a reasonable model. As has been pointed out
above, the validity of this process can be checked for the case of
cosmic strings, where a more rigorous derivation has been done.

Let us  start by defining a characteristic length scale
\be
L^{4-n}=\frac{M}{\rho}\,, \label{charscale}
\ee
where $M$ will have dimensions appropriate for the defect in question (i.e., monopole mass, string mass per unit length, or wall mass per unit area), and can also be written
\be
M\sim \eta^n, \label{massscale}
\ee
with $\eta$ being the symmetry breaking scale. Also, we interpret the velocity as being the RMS velocity of the defect network, and allow for energy losses due to interactions, which for extended defects can usually be modeled (purely on dimensional grounds) by
\be
\frac{d\rho}{dt}=-c\frac{v}{L}\rho\,; \label{energyloss}
\ee
we shall show below that this $\rho$ dependence also applies to global
monopoles while for local monopoles the energy loss is proportional to
$\rho^2$.  More importantly, defects will be slowed down by friction
due to particle scattering. This can be characterized by a friction
length scale \be {\bf f}=-\frac{M}{\ell_f}\gamma {\bf v}
\label{fricscale} \ee where we are defining \be \ell_f\equiv
\frac{M}{\theta T^{n+1}}\propto a^{n+1} \ee and $\theta$ is a parameter counting the number of particle
species (or degrees of freedom) interacting with the defect.

We can also define an overall damping length which includes both the effect of Hubble damping and that friction due to particle scattering
\be
\frac{1}{\ell_d}=nH+\frac{1}{\ell_f}\,.
\ee
It is important to compare the relative importance of the two effects. Since the friction length scale will in most circumstances grow faster than the Hubble length, it is expected that friction will be dominant at early times, while Hubble damping will dominate at sufficiently late times. This is easy to confirm. Using the relations \cite{KOLB}
\be
t\sim\frac{m_{Pl}}{T^2}
\ee
for the radiation era, and
\be
t\sim\frac{m_{Pl}}{T^{3/2}T_{eq}^{1/2}}
\ee
for the matter era, we find that the ratio of the Hubble and friction length scale is given by
\be
\frac{1}{\ell_f H}\sim\left(\frac{m_{Pl}}{\eta}\right)^n \left(\frac{T}{m_{Pl}}\right)^{n-1}
\ee
in the radiation era, and
\be
\frac{1}{\ell_f H}\sim\left(\frac{m_{Pl}}{\eta}\right)^n \left(\frac{T}{m_{Pl}}\right)^{n-1/2} \left(\frac{m_{Pl}}{T_{eq}}\right)^{1/2}
\ee
in the matter era, where again $\eta$ is the scale of the phase transition producing the defects. In the interest of simplicity, these estimates neglect couplings as well as numbers of interacting species (which may not necessarily be of order unity). For defects formed at the GUT phase transition, the friction length scale is $\sim 10^3$ times smaller than the Hubble length. Note that the ratio is constant for monopoles in the radiation era - hence monopoles will always be friction-dominated in the radiation era (but not in the matter era). Apart from this special case, friction domination ends at
\be
\frac{T}{m_{Pl}}\sim\left(\frac{\eta}{m_{Pl}}\right)^{\frac{n}{n-1}}
\ee
if in the radiation era, otherwise at
\be
\frac{T}{m_{Pl}}\sim\left(\frac{\eta}{m_{Pl}}\right)^{\frac{2n}{2n-1}} \left(\frac{T_{eq}}{m_{Pl}}\right)^{\frac{1}{2n-1}}
\ee
in the matter era.

Putting together all of the above effects, we find the following evolution equation for the characteristic length scale $L$ and RMS velocity $v$
\be
(4-n)\frac{dL}{dt}=(4-n)HL+v^2\frac{L}{\ell_d}+cv
\ee
\be
\frac{dv}{dt}=(1-v^2)\left(f-\frac{v}{\ell_d}\right)
\ee
where in the latter we have included the possibility of further driving forces affecting the defect dynamics. Note that $f$ has the units of acceleration, ie it is the force per unit mass. For extended objects (walls and strings) that have been extensively studied in the past, this driving force is obviously the local curvature, and we have
\be
f\sim\frac{k}{L}\,;
\ee
we are implicitly assuming that our characteristic length scale is the same as the defect curvature radius. For monopoles the situation is more complicated, since there are forces due to other monopoles. This will be discussed below.

%%%%%%%%%%%%%%%%%%%%%%%%%%%%%%%%%%%%%%%%%%%%%%%%%%%%%%%%%%%%%%%%%%%%%%%%%%
\section{\label{monop}Overview of Monopoles}

We now provide a brief overview the properties of local and global monopoles. This is by no means exhaustive---we shall only focus on the dynamical properties that will turn out to be relevant for our analytic model. We shall follow the more extensive reviews in \cite{LECTURESC,LECTURESP,VSH}, and refer the interested reader to them for a more detailed discussion.

The local monopole solution has two characteristic length scales, $r_s$ and $r_v$, identifying the radii of regions in which the scalar and vector fields depart significantly from their asymptotic behaviour. The monopole mass is approximately given by
\be
m\sim\frac{4\pi}{e^2}m_v\sim\frac{4\pi}{e}\eta_m.
\ee
Their initial separation, $\xi_i$, can range from the monopole thickness (for a second-order phase transition) to the horizon size (for a strongly first-order phase transition). The expansion of the universe will dilute the monopoles, so naively we might expect a solution of the form $\xi\propto T^{-1}\propto a$.

The force between local monopoles is just electromagnetic attraction,
\be F_{emg}(r)\sim\frac{g^2}{r^2} \ee where the magnetic charge is
given by $g= 4\pi/e$. A monopole moving through a plasma also
experiences a drag force due to its interaction with charged
particles; the corresponding force is \be {\bf F}_{plasma}\sim - \theta
T^2 \gamma{\bf v} \ee in agreement with Eqn. (\ref{fricscale}). The
initial monopole density is $n_M\propto\xi^{-3}$, though note that due
to the Coulomb forces the positions of monopoles and antimonopoles are
strongly anti-correlated: except, possibly, for a short transient,
the nearest neighbour to a monopole is likely to be an anti-monopole.
The Kibble mechanism would naturally produce such an
  anticorrelation already at formation. On the other hand, for slow
  transitions there is a competing mechanism in which the monopoles
  are the result of thermal fluctuations of the gauge field
  \cite{Hindmarsh:2000kd,ARTTU} and they form in same-charge clusters.
  However, even in this case the typical number of monopoles per
  cluster is of order unity at the relevant energy scales around the
  GUT scale \cite{ARTTU}.  Then the attractive forces between
  monopoles and antimonopoles will dynamically establish the
  anti-correlations very quickly.

One might naively expect monopole motion in a plasma to be like
Brownian motion of heavy dust particles in a gas or liquid. If this is
so, then they would typically move with thermal velocities \be
v_T\sim\left(\frac{T}{m}\right)^{1/2} \label{thermal1} \ee with
mean-free path \be \ell\sim\frac{1}{\theta
  T}\left(\frac{m}{T}\right)^{1/2}\,. \label{thermal2} \ee However,
the Coulomb forces will again introduce a bias in the random walks,
and so the defects gradually drift towards each other. The average
drift velocity may be simply estimated from the balance between the
electromagnetic and drag forces, leading to \be v\sim\frac{g^2}{\theta
  T^2r^2}\,.  \ee This dissipates some energy and leads to the
formation of bound monopole-antimonopole states, at a capture radius
\be r_c\sim\frac{g^2}{T}\,.  \ee

Once the bound pair is formed, the monopole and antimonopole spiral in, losing energy to the plasma drag and radiation, and eventually annihilate into gauge bosons. The rate of radiative energy loss into gauge bosons can be estimated using the classical electromagnetism radiation formula
\be
{\dot\E}\sim-\frac{(ga)^2}{6\pi}
\ee
($a$ being the monopole acceleration), which leads to
\be
{\dot\E}_{gauge}\sim-\frac{g^6}{m^2r^4}\sim-\frac{g^4}{\eta_m^2r^4}\,. \label{localloss}
\ee
The lifetime of the bound state is
\be
\tau\sim\frac{m^2r^3}{g^4}\sim\frac{\eta_m^2r^3}{g^2}\,.
\ee
Notice that radiation losses increase strongly as the separation decreases, and indeed almost all the energy is lost after capture. 

These bound states are generically short-lived, although if desired one could design models where the annihilation would occur much later than the capture. In any case, this subtlety can in practice be neglected, since a captured pair is doomed to annihilate anyway---it can therefore be considered as decoupled from the rest of the network from the moment of capture onwards.

This, together with the fact that most radiation losses occur in bound pairs, means that we need not model these effects explicitly. All we require is an adequate account of the `pair losses', that is we need to keep track of the rate at which bound pairs are forming and leaving the `free' monopole network. This is analogous to neglecting the effect of the small string loops when modeling the dynamics of a network of `infinite' strings---there one also does not need to explicitly account for losses to gravitational radiation (which occur mostly in the loops rather than in the infinite strings).

Finally, note that the diffusive capture is only effective if the monopole mean free path in the plasma is smaller than the capture radius, which corresponds to
\be
T>\frac{m}{\theta^2g^4}\sim\frac{\eta_m}{\theta^2g^3}\,.
\ee
At lower temperatures the monopoles and antimonopoles can only capture each other by emitting radiation. For thermal incident velocities we have already seen that the energy losses are very small, so the annihilation time is much larger than the expansion rate and the monopole to entropy ratio becomes effectively frozen.

More interesting still is the case of global monopoles. The total
energy of a monopole out to a given distance grows linearly with
distance, so in some sense global monopoles are similar to local
strings. The energy of a monopole-antimonopole pair at a distance $R$
is \be \E\sim4\pi\eta^2 R\,,\label{geq1} \ee and the corresponding
attractive force is therefore \be F_{global}\sim\frac{\partial
  \E}{\partial R}\sim 4\pi\eta^2 \ee which is independent of
distance. This is analogous to monopoles connected by strings.
But unlike that case, a global monopole is not paired with any
particular antimonopole, and so it is not \textit{a priori} clear how
efficiently it can find a partner. Still, the existence of long-range
forces means that capture and annihilation will be more efficient than
in the local case, and therefore no `monopole problem' is expected
here. The rate of energy loss of a monopole-antimonopole bound state
into Goldstone bosons has been estimated to
be  \cite{Barriola} \be
{\dot\E}_{gold}\sim-\eta^2 \label{globalloss} \ee and its
lifetime is $\tau\sim R$. The pair is expected to move at
ultra-relativistic speeds and annihilate within a Hubble time. Again,
this allows us to model the energy losses as going into bound pairs as
opposed to modeling radiation losses explicitly.

Finally, let us note that there is some previous numerical work on global monopole networks. Numerical simulations in the non-linear sigma model approximation by Bennett and Rhie \cite{BRHIE} suggest that a scaling solution is reached with a few monopoles per horizon volume, namely
\be
n_Hd^3_H=3.5\pm1.5
\ee
in the radiation era and
\be
n_Hd^3_H=4.0\pm1.5
\ee
in the matter era. More recently, Yamaguchi \cite{YAMAGUCHI} gives for the quantity $\xi=nt^3$,
\be
\xi_r=0.43\pm 0.07
\ee
\be
\xi_m=0.25\pm0.05
\ee
and also attempts to measure velocities, finding
\be
v_r=1.0\pm0.3
\ee
\be
v_m=0.8\pm0.3\,.
\ee
Note that for a scale factor $a(t)\propto t^\lambda$
\be
d_H(t)=\frac{t}{1-\lambda}
\ee
so the two sets of simulations agree with each other remarkably well in the radiation era, and broadly so in the matter era.

%%%%%%%%%%%%%%%%%%%%%%%%%%%%%%%%%%%%%%%%%%%%%%%%%%%%%%%%%%%%%%%%%%%%%%%%%%
\section{\label{modelspec}Modeling Monopoles}

We are now in a position to bring together the results of the two previous sections and discuss how the velocity-dependent one-scale model can be extended to describe the evolution of monopole networks. From the above discussion, it is clear that the force between a pair of local monopoles has the following form
\be
f_{local}\sim\frac{k}{\eta L^2}\,.
\ee
For global monopoles the situation is slightly more subtle, but also more interesting. The force between a pair of them is independent of distance, but recall that their mass grows proportionally to the distance. Therefore the acceleration is in fact inversely proportional to distance,
\be
f_{global}\sim \frac{k}{L}\,.
\ee
Again we see that they are in some sense like local strings. This is not too difficult to understand: a monopole will be effectively heavier when seen on larger scales, and its acceleration should therefore be correspondingly smaller.

The next issue is the fact that there can be many monopoles and anti-monopoles in a given Hubble volume, so the various forces acting on a given one will partially cancel each other. A simple way in which one can try to model this as a $1/\sqrt{N}$ effect. In other words, the acceleration $f$ becomes $f/\sqrt{N}$, where the number of defects $N$ in a Hubble volume $d_H^3$ is given by
\be
N_g=\left(\frac{d_H}{L}\right)^3\,.
\ee
This should be fine for global monopoles (although it is an issue that warrants testing in numerical simulations). On the other hand, in the local case the situation is again more subtle due to the existence of anti-correlations in the positions of monopoles and anti-monopoles. The analysis of \cite{EINHORN} indicates that the number of defects in that case is instead approximately given by
\be
N_l\sim\left(\frac{d_H}{L}\right)^2\,.
\ee
This is to be expected: since the nearest neighbour to a monopole is likely to be an antimonopole (and vice-versa), then typically the attractive forces between neighbouring pairs will be larger than in the uncorrelated case. In other words, the cancellation mechanism is less strong, which is equivalent to saying that the effective number of neighbours is smaller.

Finally, there is the issue of energy losses due to monopole
annihilations. The generic form given by Eqn. (\ref{energyloss}), a
natural extension of those that can be derived for strings and domain
walls, is valid for the case of global monopoles (again, these are in
some sense like local strings). For a single monopole-antimonopole
pair ${\dot\rho}/\rho\sim {\dot\E}/\E$, and Eqns. (\ref{geq1}) and
(\ref{globalloss}) lead to ${\dot\rho}\propto-\rho/R$. This is another
way of saying that the timescale for energy losses corresponds (in the
fundamental units we are using) to the lenghtscale $R$. On the
assumption that the two lengthscales are comparable $R\sim L$ and this
matches Eqn. (\ref{energyloss}) apart from the allowance for generic
velocities. Note that here we are assuming that the separation between
monopole-antimonopole pairs is comparable to the network's
characteristic lengthscale. Although the two need not be the same,
this is a valid assumption in the context of the simple one-scale
model we are considering, and additionally any discrepancy could to
some extent be absorbed by a redefinition of the numerical coefficient
$c$.

On the other hand, in the local case the Coulomb forces between the
monopoles and antimonopoles lead to a different energy loss rate.
Early work of Zel'dovich and Khlopov \cite{KHLOPOV} assuming that the
monopole-antimonopole annihilations were purely determined by the
Coulomb attraction of the magnetic charges found that the number
density $n_M$ should evolve as 
\be \frac{dn_M}{dt}+3Hn_M=-A\frac{n_M^2}{T^3}; \ee 
a subsequent more detailed study by Preskill \cite{PRESKILL}, which
implicitly allows for the effect of the anti-correlations, leads to
the more generic form \be \frac{dn_M}{dt}+3Hn_M=-A\frac{n_M^2}{T^p}, \ee
where the proportionality constant has the form $A=C\eta^{p-2}$, and
$C$ is a dimensionless constant.

On physical grounds we should expect that $p\le3$. In terms of the correlation length this has the form
\be
3\frac{dL}{dt}=3HL+\frac{A}{L^2T^p}\,.\label{preskiloss}
\ee
This may seem very different from the standard term given by Eqn. (\ref{energyloss}), but the difference is actually less than it appears. For example, there should be a velocity dependence in this term (for the obvious reason that if all monopoles have zero velocities there will be no annihilations). This is explicit in the standard term, but implicit in the Preskill term. An easy way to see it is to assume that monopoles have thermal velocities. In this case one can use Eqns. (\ref{thermal1}-\ref{thermal2}) to put in a phenomenological but explicit velocity dependence. We then find
\be
\frac{A}{L^2T^p}\sim Cv\left(\frac{\eta}{T}\right)^{p-5/2}
\ee
though we caution that the above assumption of thermal velocities is, at best, a crude approximation. In any case, we will use the Preskill loss term, Eqn. (\ref{preskiloss}), for the local monopoles, and comment of the differences we would obtain otherwise.

As for the specific values of $p$, Preskill argues for a short, transient high-temperature regime where $p=2$, and a longer low-temperature regime with $p=9/10$. This is important because if $p<1$ annihilations are expected to 'turn off' (that is, become unimportant), in which case we expect $n\propto T^3$, which corresponds to $L\propto t^{1/2}$. On the other hand, if $p>1$ then annihilations are always relevant, and in that case one expects $n\propto T^{p+2}$, which corresponds to $L\propto t^{(p+2)/6}$. (Notice that this analysis is for the radiation dominated epoch.) It is therefore important to see if we can recover these results using the model in the local case. For the global case, a good benchmark will be the simulations discussed above.

%%%%%%%%%%%%%%%%%%%%%%%%%%%%%%%%%%%%%%%%%%%%%%%%%%%%%%%%%%%%%%%%%%%%%%%%%%
\section{\label{local}Evolution of Local Monopoles}

We can now use our newly derived analytic model equations to look for scaling solutions for the characteristic lengthscale and RMS velocity of the monopoles. Starting with the case of local monopoles, the evolution equations have the general form
\be
3\frac{dL}{dt}=3HL+v^2\frac{L}{\ell_d}+\frac{C\eta^{p-2}}{L^2T^p}
\ee
and
\be
\frac{dv}{dt}=(1-v^2)\left(\frac{k}{\eta L^2}\frac{L}{d_H}-\frac{v}{\ell_d}\right)\,.
\ee

We can start by finding solutions in Minkowski space-time, by setting $H=0$. In this case the asymptotic scaling solution has the form
\be
L^3\propto\frac{\eta^{p-2}}{T^p} t \label{localL}
\ee
and the monopoles will freeze, with the precise scaling law for the velocity depending on the behavior of the friction length scale. Assuming that we have $\ell_f=const$ we find
\be
v\propto t^{-4/3}, \label{localvf}
\ee
while for the arguably more realistic $\ell_f\propto L$ the freezing happens more slowly,
\be
v\propto t^{-1}\,. \label{localvl}
\ee
In the (unrealistic) frictionless limit $\ell_f\to\infty$ the correlation length still has the same scaling, but velocities asymptote to the speed of light, $v\to 1$.

For the expanding case, allowing for a generic scale factor of the form $a\propto t^{\lambda}$, there are two possible scaling laws, which depend both on the value of $p$ and on $\lambda$. For the case
\be
p<3-\frac{1}{\lambda}
\ee
we will have
\be
L\propto t^{\lambda}\propto a; \label{locall1}
\ee
this corresponds to the case where energy losses due to annihilation are unimportant and the monopoles are simply conformally stretched. Note that in the radiation era we do recover the result $L\propto t^{1/2}$, and in that case the threshold for this regime is indeed $p<1$. This therefore recovers and generalizes the Preskill results. In the opposite case
\be
p>3-\frac{1}{\lambda}
\ee
then annihilations are dynamically important, and the scaling law for the correlation length is then
\be
L\propto t^{(\lambda p+1)/3}; \label{locall2}
\ee
again we recover the expected Preskill result $L\propto t^{(p+2)/6}$ for the particular case of the radiation era. Note that for $p>3-1/\lambda$ we have $(\lambda p+1)/3>\lambda$: as expected, in this regime the evolution is faster than the one above, which corresponds to conformal stretching. This difference illustrates the effect of the annihilations.

It is also worth pointing out that linear scaling ($L\propto t$) will
occur for the case $\lambda p=2$, while for $\lambda p>2$ the
correlation length will grow superluminally. This last case
corresponds (at a phenomenological level) to the situation where
annihilations are so efficient that on average there will eventually
be less than one monopole per Hubble volume, so the monopoles
effectively disappear. Note that since we physically expect $p\le3$,
then $\lambda p=2$ corresponds to $\lambda\ge2/3$. Hence linear
scaling can occur in the matter era but not in the radiation era. On
the other hand, superluminal scaling requires $\lambda>2/3$ and so it
can't occur in either epoch---this is a simple manifestation of the
monopole problem in standard cosmology.

Interestingly, in both regimes the scaling law for the velocities is the same, namely
\be
v\propto t^{-\lambda}\propto a^{-1}\propto T\,. \label{localvv}
\ee
This is a nice and simple result, and it disproves the naive expectation that the monopoles should move with thermal velocities (which would correspond to $v\propto \sqrt{T}$). 

The above solutions hold for decelerating universes (with $0<\lambda<1$) but also for power-law inflating universes (that is, with $\lambda\ge1$). On the other hand, in de Sitter space (with $a\propto e^{Ht}$) we have
\be
L\propto a,\quad p\le3
\ee
or
\be
L\propto a^{p/3},\quad p>3\,,
\ee
although we expect that the latter behavior for $p$ is physically unrealistic. For the velocity we still have
\be
v\propto a^{-1}\,.
\ee
Thus in an inflating universe the monopoles freeze and are conformally stretched, being pushed outside the horizon. After the end of inflation there will be much less than one monopole per horizon and their velocities will be infinitesimal, so they will keep being conformally stretched until they re-enter the horizon. So in order to solve the monopole problem one needs sufficient e-folds of inflation to ensure that the monopoles have not yet re-entered.

%%%%%%%%%%%%%%%%%%%%%%%%%%%%%%%%%%%%%%%%%%%%%%%%%%%%%%%%%%%%%%%%%%%%%%%%%%
\section{\label{global}Evolution of Global Monopoles}

A similar analysis can now be done for the global case. We shall see that the different force and energy loss terms will lead to very different scaling laws. This case is also interesting because previous numerical simulations exist against which we can compare our results. Although the simulations have been done several years ago and have very low resolution by today's standards, we shall see that the results of the comparison are very encouraging.

\subsection{Analytic Model}

In this case the evolution equations will have the general form
\be
3\frac{dL}{dt}=3HL+v^2\frac{L}{\ell_d}+cv
\ee
and
\be
\frac{dv}{dt}=(1-v^2)\left[\frac{k}{L}\left(\frac{L}{d_H}\right)^{3/2}-\frac{v}{\ell_d}\right]\,.
\ee

Again we can start with the Minkowski space-time case. In the unrealistic frictionless limit $\ell_f\to\infty$ we now have asymptotically
\be
L=\frac{1}{3}ct
\ee
\be
v=1
\ee
so global monopoles will become ultra-relativistic. The case of a constant friction length scale is not relevant for global monopoles. Due to the linear divergence of their masses, a more realistic situation in Minkowski space time would be that of the friction length scale being proportional to the correlation length itself, $\ell_f\propto L$. In that case we find the following scaling law 
\be
L\equiv\epsilon t=\frac{1}{3}v_0(v_0+c)t \label{gloflatl}
\ee
\be
v=k\epsilon^{3/2}=const. \label{gloflatv}
\ee
This behavior is to be contrasted with the case of local monopoles, whose velocities always approach zero (except in the unrealistic frictionless case). Note that in principle any value of the velocity is a possible solution, including the limit $v= 1$. An interesting question is whether the friction will make the monopole velocities stabilize at some fixed value (and if so, how small this is) or if they will still become arbitrarily close to the speed of light. Incidentally, notice that assuming $\ell_f\propto t^\sigma$ in Minkowski space, requiring a linear scaling $L\propto t$ implies $v=const$ and $\sigma=1$. In other words, no other non-trivial behavior of the friction length would lead to linear scaling for the network.

Now let us consider the general expanding case, again for a generic expansion law $a\propto t^{\lambda}$. Here, just as in the standard case of local strings, the only possible scaling law is linear scaling
\be
L=\epsilon t
\ee
\be
v=v_0=const.,
\ee
which at least qualitatively is in agreement with the existing numerical simulations. Interestingly, just as in the Minkowski case there are two branches of the solution: the velocities may or may not be ultra-relativistic! Firstly, the ultra-relativistic scaling regime will have the following scaling parameters
\be
v_0=1
\ee
\be
\epsilon=\frac{c}{3-4\lambda};
\ee
note that this can only hold for $\lambda<3/4$, but is in principle allowed both in the radiation and in the matter eras. Secondly, the more standard (sub-luminal) scaling regime will be characterized by the following scaling parameters
\be
\epsilon=\frac{cv_0}{3(1-\lambda)-\lambda v_0^2}
\ee
\be
\lambda v_0=k(1-\lambda)^{3/2}\epsilon^{1/2};\label{glovalvg}
\ee

These relations could be solved explicitly for $\epsilon$ and $v_0$, but the corresponding expressions would not be too illuminating. However, simplified and physically suggestive solutions can be displayed for both limits of the expansion power $\lambda$. In the limit $\lambda\to0$, we have 
\be
\epsilon=\frac{1}{3}cv_0\,.
\ee
Not surprisingly, this is similar to the Minkowski space-time scaling we discussed above. On the other hand, in the limit $\lambda\to1$, we find
\be
v_0=\frac{1}{3}ck^2(1-\lambda)^2\,.
\ee
Here the scaling velocity becomes arbitrarily small ($v\to0$) and $L\propto a\propto t$ so asymptotically this is a conformal stretching regime.

Unlike the ultra-relativistic branch, this non-luminal branch can exist for any $\lambda$ (that is, for any expansion law), though note that there is a constraint on the scaling value of the velocity
\be
v_0^2<3(\frac{1}{\lambda}-1);
\ee
this is trivial for $\lambda<3/4$ (meaning that in such cases any scaling value for the velocity is allowed in principle), but restrictive for faster expansion rates. Note in particular that it agrees with the above finding that $v=1$ is only allowed for $\lambda<3/4$.

In passing we also note that the linear scaling solution $L\propto t$, $v=const$ will also hold if we consider the case where the friction length varies as $\ell_f\propto L$ instead of the usual scaling with temperature. Even in that case no other scaling solutions exist. The only change is that in the scaling coefficients ($\epsilon$ and $v_0$ above), we would need to interpret the parameter $\lambda$ as having a renormalized value, instead of the value given by the expansion rate.

We can also find scaling solutions in inflating universes. Here there is a unique solution, both for power-law inflation and for the de Sitter case, namely
\be
L\propto a
\ee
\be
v\propto a^{-1};
\ee
hence the solution is the same as in the local case, and the number of e-folds of inflation required to keep the monopoles outside the horizon by the present day should be the same in both cases.

\subsection{Comparing to Simulations}

Our results suggest that it would be very interesting to carry out high-resolution numerical simulations of global monopoles with a range of different expansion rates (say, radiation, matter and a value of $\lambda>3/4$) in order to check these solutions and provide a good calibration for the model. In the meantime, however, we can use the results of Bennett and Rhie \cite{BRHIE} and of Yamaguchi \cite{YAMAGUCHI} for the correlation length, and also the latter's for the velocities, in order to make some simple comparisons.

We start by translating these results into our scaling parameter $\epsilon$, finding
\be
\epsilon_r\sim1.32,\quad \epsilon_m\sim1.89
\ee
for the simulations of Bennett and Rhie, and
\be
\epsilon_r\sim1.32,\quad \epsilon_m\sim1.59
\ee
for Yamaguchi's. Notice the remarkable agreement in the radiation era; even in the matter era the difference is small considering the relatively low resolution and dynamic range of the simulations. Now, since Yamaguchi's measurement is consistent with luminal velocities, let us assume that we are in the $v=1$ branch, and solve for the energy loss parameter $c$. We then find
\be
c_r\sim1.32,\quad c_m\sim0.63
\ee
for the simulations of Bennett \& Rhie, and
\be
c_r\sim1.32,\quad c_m\sim0.53;
\ee
notice that there is a factor of two difference between the values of the parameter in the radiation and matter eras, while we would expect to find similar values in both epochs if the model is broadly correct and the parameter $c$ is a constant (or nearly so). On the other hand, if we assume that we are in the sub-luminal branch (using in both cases Yamaguchi's values for the velocities, since Bennett and Rhie doesn't provide a measurement),
\be
c_r\sim1.32,\quad c_m\sim1.35
\ee
for the simulations of Bennett and Rhie, and
\be
c_r\sim1.32,\quad c_m\sim1.15;
\ee
here we can claim a very good agreement, considering the resolution of the simulations in question and the error bars that each of them has. Also in this branch we can compare the scaling values of the velocities; according to the model we expect the ratio between the matter and radiation era scaling values to be
\be
\frac{v_m}{v_r}=\left(\frac{\epsilon_m}{6\epsilon_r}\right)\sim0.5,
\ee
while Yamaguchi finds $v_m/v_r\sim0.8$; again given the error bars we would argue that the agreement is encouraging.

%%%%%%%%%%%%%%%%%%%%%%%%%%%%%%%%%%%%%%%%%%%%%%%%%%%%%%%%%%%%%%%%%%%%%%%%%%
\section{\label{whatif}Revisiting our assumptions}

As we discussed in Sect. \ref{modelspec}, most of the subtlety in the analytic description of these defect networks rests in the way the energy losses, the inter-monopole forces and their suppression factors are incorporated into the model. In this section we will revisit these assumptions and discuss the extent to which they influence the results we presented so far.

Let us start by commenting on the role of the annihilation term, by considering what would happen if we had used Preskill's annihilation term---Eqn. (\ref{preskiloss}), which holds for local monopoles---for the global monopole analysis of Sect. \ref{global}. For scaling laws of the form $L\propto t^\alpha$, $v\propto t^\beta$, we find the following scaling powers for the correlation length
\be
\alpha=\lambda,\quad  p<3-\frac{1}{\lambda}
\ee
and
\be
\alpha=\frac{\lambda p +1}{3},\quad  p>3-\frac{1}{\lambda}
\ee
(again annihilations are unimportant in the former case but not in the latter). Not surprisingly the scaling laws for $L$ would be similar to those we found in the local case (Sect. \ref{local}). On the other hand, for the velocities we have
\be
\beta=-\lambda,\quad  \lambda<1/3
\ee
and
\be
\beta=\frac{\alpha-1}{2},\quad  \lambda>1/3.
\ee
Again the first of the above existed in the local case (for all values of $\lambda$), while the second branch is new. So there would in general be no linear scaling solution; indeed the only possibility of having $\alpha=1$ is for the special case $\lambda p=2$ (with $p>2$). But in any case the velocities will become arbitrarily small, so as expected this form of annihilation term seems clearly ruled out even by the existing, low-resolution numerical simulations.

A more important issue is the suppression factor on the forces driving the monopoles, to account for the partial cancellation due to the presence of many monopoles and antimonopoles in a given Hubble volume. In Sect. \ref{modelspec} we argued that this could be described as a $1/\sqrt{N}$ effect, but with the number $N$ of defects in a Hubble volume calculated differently for the local and global cases. Here we will generalize this assumption, by assuming that $N$ is generically given by
\be
N_s\sim\left(\frac{d_H}{L}\right)^{3-s}\,;
\ee
recall that in the above analysis it was assumed that $s=0$ for the global case but $s=1$ for the local case.

Our analysis in Sects. \ref{local} and \ref{global} can now be repeated for this generic suppression term. The general outcome is that our previous analysis is fairly robust. Most scaling laws are not affected or receive $s$-dependent corrections only in the pre-factors. Corrections in the scaling exponents appear only for the velocities, not the correlation lengths. On the other hand, the emergence of genuinely new scaling regimes is only possible for (arguably) unrealistic values of the parameter $s$. We now discuss these changes separately for the local and global cases.

\subsection{Local case}

The evolution equations now take the form \be
3\frac{dL}{dt}=3HL+v^2\frac{L}{\ell_d}+\frac{C\eta^{p-2}}{L^2T^p} \ee
and \be \frac{dv}{dt}=(1-v^2)\left[\frac{k}{\eta
    L^2}\left(\frac{L}{d_H}\right)^{(3-s)/2}-\frac{v}{\ell_d}\right]\,.
\ee Recall that our expectation, based on the results of Einhorn
\textit{et al.} \cite{EINHORN} is that $s=1$: the suppression factor
describing the partial cancellation of the force term is weaker due to
monopole-antimonopole anticorrelations, which can be reinterpreted
as an reduction in the effective number of neighbours.

Starting with Minkowski case ($H=0$), there are no changes in the asymptotic case $\ell_f\to\infty$. For $\ell_f=const$ the scaling law for the lengthscale $L$ (given in Eqn. (\ref{localL})) also remains unchanged, for any values $s\le7/2$, On the other hand, the power law of the scaling of the velocities ($v\propto t^\beta$) does have an $s$-dependence
\be
\beta=\frac{s-5}{3}\,;
\ee
for $s=1$ we therefore recover our previous result in Eqn. (\ref{localvf}), $\beta=-4/3$. Something analogous happens for the arguably more realistic case $\ell_f\propto L$: the scaling law for $L$ is unchanged for $s\le3$, while the power law of the velocities becomes
\be
\beta=\frac{s-4}{3}\,;
\ee
again for $s=1$ we recover Eqn. (\ref{localvl}), $\beta=-1$. In this case there is in principle a new scaling branch for $s\ge3$ (even though we emphasize that such a suppression factor is physically unrealistic). In this case we would have $\beta=-1/s$ and the characteristic lengthscale would also scale differently: $L\propto t^\alpha$ with $\alpha=1-(2/s)$.

For the case of the expanding universe, we again find that the scaling law for the lengthscale remains unchanged, being given by Eqns. (\ref{locall1}) or (\ref{locall2}) as before, depending on the value of the Preskill factor $p$. For the velocities there are two possibilities. If $s\le1$ the scaling law is again unchanged, being given by $\beta=-\lambda$ (for $a\propto t^\lambda$) as in Eqn. (\ref{localvv}). On the other hand, for $s>1$ the velocities will scale differently, with the scaling exponent being given by
\be
\beta=-\frac{\alpha}{2}(s+1)+\frac{1}{2}(s-1)\,,
\ee
where again $\alpha$ is the scaling exponent for $L$ given by Eqns. (\ref{locall1}) or (\ref{locall2}). Careful measurements of the scaling laws for $L$ and $v$ in numerical simulations with several expansion rates can therefore be used to not only test the analytic model but also to obtain indirect information on the inter-monopole forces.

\subsection{Global case}

In this case the evolution equations will have the generic form
\be
3\frac{dL}{dt}=3HL+v^2\frac{L}{\ell_d}+cv
\ee
and
\be
\frac{dv}{dt}=(1-v^2)\left[\frac{k}{L}\left(\frac{L}{d_H}\right)^{(3-s)/2}-\frac{v}{\ell_d}\right]\,,
\ee
where thus far we have assumed $s=0$, corresponding to the expectations that in the absence of any correlation or anticorrelations the suppression factor should be a simple $1/\sqrt{N}$ effect.

Again starting with the Minkowski space case ($H=0$), the case $\ell_f\to\infty$ is still unchanged. For the case $\ell_f\propto L$ the linear scaling solution $L=\epsilon t$, $v=const$ given by Eqns. (\ref{gloflatl}) and (\ref{gloflatv}) still exists for all values of $s$, the only difference being that the relation between the pre-factors becomes $s$-dependent
\be
v=k\epsilon^{(3-s)/2}\,,
\ee
which trivially reduces to Eqn. (\ref{gloflatv}) for $s=0$. (In particular, monopole velocities arbitrarily close to the speed of light are still possible in principle.) Interestingly, for the case $s=1$ only, there is a second scaling solution: writing $L\propto t^\alpha$ and $v\propto t^\beta$ as usual, this solution is characterized by
\be
\alpha=1+\beta=\frac{1}{3}ck\,,
\ee
subject to the constraints that $\alpha<1$ and $\beta<0$ which trivially imply $ck<3$. Unlike the generic linear scaling regime, here the characteristic lengthscale grows more slowly and the velocities become arbitrarily small instead of being constant.

Something similar occurs in the case of the expanding universe. The linear scaling solution still exists for all values of $s$ (including the branch with $v=1$), the only difference being an $s$-dependence in the relation between the pre-factors for the $L$ and $v$ scaling laws, formerly given by Eqn. (\ref{glovalvg}) and now becoming
\be
\lambda v_0=k(1-\lambda)^{(3-s)/2}\epsilon^{(1-s)/2}\,,
\ee
which trivially reduces to the previous result when $s=0$.

In this case a second, non-relativistic scaling solution appears only for the case $s=-1$. The physical viability (if any) and interpretation of such a term is somewhat unclear: this would imply that the suppression factor is stronger than the naive $1/\sqrt{N}$ effect, or in other words each monopole has a larger effective number of neighbours. Just as in the Minkowski case, the new scaling solution is characterized by
\be
\alpha=1+\beta\,,
\ee
subject to $\alpha<1$ and $\beta<0$, but the specific values of the scaling coefficients now depend on the expansion rate, being given by
\be
(\beta+\lambda)(1+\beta-\lambda)=\frac{1}{3}(1-\lambda)^2ck\,
\ee
and subject to the further constraints that both terms in brackets on the left-hand side of the above equation are positive. 

Although this scaling solution is clearly related to the one we discussed previously in Minkowski space, one can't recover the Minkowski one simply by taking the limit $\lambda\to0$: this is due to the fact that the behavior of the damping terms is different in the two cases ($\ell_f\propto L$ for the Minkowski case versus Hubble damping for the expanding case). This is also the reason why different values of $s$ are needed in the two cases for the solution to exist. The two new scaling solutions have the distinguishing feature of velocities decreasing as the network evolves and becoming arbitrarily small (as opposed to being constant), which again highlights the need for careful measurements of monopole velocities.

%%%%%%%%%%%%%%%%%%%%%%%%%%%%%%%%%%%%%%%%%%%%%%%%%%%%%%%%%%%%%%%%%%%%%%%%%%
\section{\label{concl}Summary and outlook}

We have proposed a velocity-dependent one-scale phenomenological
description of the evolution of a network of pointlike defects
--monopoles--, both in the global and local (magnetic) cases.  In both
cases we recover scaling solutions previously found in the literature
and make predictions for new scaling regimes that have not been
considered before.

In the global monopole case there are three main assumptions.  First,
since the force between global monopoles is approximately independent
of distance, a monopole interacts with every other monopole and
antimonopole within its Hubble volume and this reduces the effective
force it feels. Second, since its mass grows linearly with distance
the acceleration has the same $1/L(t)$ dependence as for local cosmic
strings. Finally, since radiation losses occur mainly within bound
monopole-antimonopole pairs, and are therefore decoupled from the rest
of the network in a first approximation, they can be modelled by a
single energy-loss parameter decribing the efficiency of pair
formation.  Comparison with existing numerical simulations
\cite{BRHIE,YAMAGUCHI} supports this overall picture and we recover
scaling solutions with $v<1$ previously reported in the literature.
The rms velocity of the monopoles is not determined by the model. On the
other hand, the expected $v=1$ scaling solution does not agree so well
with the existing simulations.  However the calculation of monopole
velocities from numerical simulations has large uncertainties so it
would be important to revisit this point numerically.

In the case of magnetic monopoles we build on Preskill's results
\cite{PRESKILL} for the annihilation rate.  The suppression of the
Coulomb force between monopoles due to the presence of many neighbours
is further corrected to account for anticorrelations (a monopole's
neighbour is more likely to be an antimonopole). 

Mindful of the fact that our modeling
relies on a few key assumptions, we have also discussed how our
results change if some of these assumptions are relaxed. We have found
that our scaling solutions are usually quite robust, though more
so for the defect correlation lengths than for their velocities.
In particular there can be changes if the suppression factors for the
forces driving the monopoles are different. A context where this may
possibly happen is that of condensed matter systems such as Helium or
liquid crystals. The behavior of monopole-like defects in these
systems deserves further study.

It would be very interesting to extend this analysis to the case of
hybrid semilocal networks (of monopoles connected by strings), for
which ref. \cite{Achucarro:2007sp} found some evidence of scaling.
Numerical simulation is computationally very demanding for these
systems, because one cannot rely on any thin string or sigma model
approximations.

In semilocal networks, the long strings and loops behave like local
cosmic strings but string segments are also possible and they can
dominate the dynamics (see \cite{Achucarro:1999it} for a review). At
formation the network consists only of segments
\cite{Achucarro:1998ux}, with an exponential distribution in length.
But the ends of segments have long-range interactions that make them
behave much like the global monopoles studied here. Once formed, they
will interact and annihilate with other segment ends even at long
distances, causing some segments to grow into long strings while
others will collapse and disappear.

This long-range interaction provides a crucial difference with the
better studied case of hybrid networks of monopoles connected by
strings in which the strings confine the magnetic flux of the
monopoles\cite{VSH}.  In these networks the ends of the strings are
very light and the dynamics is dominated by the tendency of the
segments to collapse.  In the semilocal case, which of these two
effects dominates the network evolution is controlled by the ratio of
the scalar and gauge couplings (see also \cite{Urrestilla:2001dd} for
a study of the electroweak case).  For large scalar quartic coupling
the strings disappear, leaving behind a scaling network of
texture-like structures. But if the coupling is small the simulations
hint at a scaling network made of local strings plus a few open
segments per horizon volume. If such structures are indeed formed
after e.g.  brane inflation, an extension of the present study would
be the right tool to confirm or rule out scaling behavior.

%%%%%%%%%%%%%%%%%%%%%%%%%%%%%%%%%%%%%%%%%%%%%%%%%%%%%%%%%%%
\begin{acknowledgments}
  We are grateful to P. Salmi and J. Urrestilla for many discussions
  in the early stages of this work. The work of C.M.  is funded by a
  Ci\^encia2007 Research Contract.  A.A.'s work is supported by the
  Netherlands Organization for Scientific Research (N.W.O) under the
  VICI programme, and by the spanish government through the
  Consolider-Ingenio 2010 Programme CPAN (CSD2007-00042) and project
  FPA 2005-04823.
\end{acknowledgments}

\bibliography{monopoles}

\begin{thebibliography}{43}
\expandafter\ifx\csname natexlab\endcsname\relax\def\natexlab#1{#1}\fi
\expandafter\ifx\csname bibnamefont\endcsname\relax
  \def\bibnamefont#1{#1}\fi
\expandafter\ifx\csname bibfnamefont\endcsname\relax
  \def\bibfnamefont#1{#1}\fi
\expandafter\ifx\csname citenamefont\endcsname\relax
  \def\citenamefont#1{#1}\fi
\expandafter\ifx\csname url\endcsname\relax
  \def\url#1{\texttt{#1}}\fi
\expandafter\ifx\csname urlprefix\endcsname\relax\def\urlprefix{URL }\fi
\providecommand{\bibinfo}[2]{#2}
\providecommand{\eprint}[2][]{\url{#2}}

\bibitem[{\citenamefont{Kibble}(1976)}]{KIBBLE}
\bibinfo{author}{\bibfnamefont{T.~W.~B.} \bibnamefont{Kibble}},
  \bibinfo{journal}{J. Phys.} \textbf{\bibinfo{volume}{A9}},
  \bibinfo{pages}{1387} (\bibinfo{year}{1976}).

\bibitem[{\citenamefont{Vilenkin and Shellard}(1994)}]{VSH}
\bibinfo{author}{\bibfnamefont{A.}~\bibnamefont{Vilenkin}} \bibnamefont{and}
  \bibinfo{author}{\bibfnamefont{E.~P.~S.} \bibnamefont{Shellard}},
  \emph{\bibinfo{title}{Cosmic Strings and other Topological Defects}}
  (\bibinfo{publisher}{Cambridge University Press},
  \bibinfo{address}{Cambridge, U.K.}, \bibinfo{year}{1994}).

\bibitem[{\citenamefont{Jeannerot}(1997)}]{Jeannerot:1997is}
\bibinfo{author}{\bibfnamefont{R.}~\bibnamefont{Jeannerot}},
  \bibinfo{journal}{Phys. Rev.} \textbf{\bibinfo{volume}{D56}},
  \bibinfo{pages}{6205} (\bibinfo{year}{1997}), \eprint{hep-ph/9706391}.

\bibitem[{\citenamefont{Jeannerot et~al.}(2003)\citenamefont{Jeannerot, Rocher,
  and Sakellariadou}}]{Jeannerot:2003qv}
\bibinfo{author}{\bibfnamefont{R.}~\bibnamefont{Jeannerot}},
  \bibinfo{author}{\bibfnamefont{J.}~\bibnamefont{Rocher}}, \bibnamefont{and}
  \bibinfo{author}{\bibfnamefont{M.}~\bibnamefont{Sakellariadou}},
  \bibinfo{journal}{Phys. Rev.} \textbf{\bibinfo{volume}{D68}},
  \bibinfo{pages}{103514} (\bibinfo{year}{2003}), \eprint{hep-ph/0308134}.

\bibitem[{\citenamefont{Sarangi and Tye}(2002)}]{Sarangi:2002yt}
\bibinfo{author}{\bibfnamefont{S.}~\bibnamefont{Sarangi}} \bibnamefont{and}
  \bibinfo{author}{\bibfnamefont{S.~H.~H.} \bibnamefont{Tye}},
  \bibinfo{journal}{Phys. Lett.} \textbf{\bibinfo{volume}{B536}},
  \bibinfo{pages}{185} (\bibinfo{year}{2002}), \eprint{hep-th/0204074}.

\bibitem[{\citenamefont{Majumdar and Christine-Davis}(2002)}]{Majumdar:2002hy}
\bibinfo{author}{\bibfnamefont{M.}~\bibnamefont{Majumdar}} \bibnamefont{and}
  \bibinfo{author}{\bibfnamefont{A.}~\bibnamefont{Christine-Davis}},
  \bibinfo{journal}{JHEP} \textbf{\bibinfo{volume}{03}}, \bibinfo{pages}{056}
  (\bibinfo{year}{2002}), \eprint{hep-th/0202148}.

\bibitem[{\citenamefont{Dasgupta et~al.}(2004)\citenamefont{Dasgupta, Hsu,
  Kallosh, Linde, and Zagermann}}]{Dasgupta:2004dw}
\bibinfo{author}{\bibfnamefont{K.}~\bibnamefont{Dasgupta}},
  \bibinfo{author}{\bibfnamefont{J.~P.} \bibnamefont{Hsu}},
  \bibinfo{author}{\bibfnamefont{R.}~\bibnamefont{Kallosh}},
  \bibinfo{author}{\bibfnamefont{A.}~\bibnamefont{Linde}}, \bibnamefont{and}
  \bibinfo{author}{\bibfnamefont{M.}~\bibnamefont{Zagermann}},
  \bibinfo{journal}{JHEP} \textbf{\bibinfo{volume}{08}}, \bibinfo{pages}{030}
  (\bibinfo{year}{2004}), \eprint{hep-th/0405247}.

\bibitem[{\citenamefont{Chen et~al.}(2005)\citenamefont{Chen, Dasgupta,
  Narayan, Shmakova, and Zagermann}}]{Chen:2005ae}
\bibinfo{author}{\bibfnamefont{P.}~\bibnamefont{Chen}},
  \bibinfo{author}{\bibfnamefont{K.}~\bibnamefont{Dasgupta}},
  \bibinfo{author}{\bibfnamefont{K.}~\bibnamefont{Narayan}},
  \bibinfo{author}{\bibfnamefont{M.}~\bibnamefont{Shmakova}}, \bibnamefont{and}
  \bibinfo{author}{\bibfnamefont{M.}~\bibnamefont{Zagermann}},
  \bibinfo{journal}{JHEP} \textbf{\bibinfo{volume}{09}}, \bibinfo{pages}{009}
  (\bibinfo{year}{2005}), \eprint{hep-th/0501185}.

\bibitem[{\citenamefont{Dasgupta et~al.}(2007)\citenamefont{Dasgupta,
  Firouzjahi, and Gwyn}}]{Dasgupta:2007ds}
\bibinfo{author}{\bibfnamefont{K.}~\bibnamefont{Dasgupta}},
  \bibinfo{author}{\bibfnamefont{H.}~\bibnamefont{Firouzjahi}},
  \bibnamefont{and} \bibinfo{author}{\bibfnamefont{R.}~\bibnamefont{Gwyn}},
  \bibinfo{journal}{JHEP} \textbf{\bibinfo{volume}{04}}, \bibinfo{pages}{093}
  (\bibinfo{year}{2007}), \eprint{hep-th/0702193}.

\bibitem[{\citenamefont{Dirac}(1931)}]{DIRAC}
\bibinfo{author}{\bibfnamefont{P.~A.~M.} \bibnamefont{Dirac}},
  \bibinfo{journal}{Proc. Roy. Soc. Lond.} \textbf{\bibinfo{volume}{A133}},
  \bibinfo{pages}{60} (\bibinfo{year}{1931}).

\bibitem[{\citenamefont{'t~Hooft}(1974)}]{THOOFT}
\bibinfo{author}{\bibfnamefont{G.}~\bibnamefont{'t~Hooft}},
  \bibinfo{journal}{Nucl. Phys.} \textbf{\bibinfo{volume}{B79}},
  \bibinfo{pages}{276} (\bibinfo{year}{1974}).

\bibitem[{\citenamefont{Polyakov}(1974)}]{POLYAK}
\bibinfo{author}{\bibfnamefont{A.~M.} \bibnamefont{Polyakov}},
  \bibinfo{journal}{JETP Lett.} \textbf{\bibinfo{volume}{20}},
  \bibinfo{pages}{194} (\bibinfo{year}{1974}).

\bibitem[{\citenamefont{Kolb and Turner}(1990)}]{KOLB}
\bibinfo{author}{\bibfnamefont{E.~W.} \bibnamefont{Kolb}} \bibnamefont{and}
  \bibinfo{author}{\bibfnamefont{M.~S.} \bibnamefont{Turner}},
  \emph{\bibinfo{title}{The Early Universe}}
  (\bibinfo{publisher}{Addison-Wesley}, \bibinfo{address}{Redwood City,
  U.S.A.}, \bibinfo{year}{1990}).

\bibitem[{\citenamefont{Goldhaber}(1989)}]{Goldhaber}
\bibinfo{author}{\bibfnamefont{A.~S.} \bibnamefont{Goldhaber}},
  \bibinfo{journal}{Phys. Rev. Lett.} \textbf{\bibinfo{volume}{63}},
  \bibinfo{pages}{2158} (\bibinfo{year}{1989}).

\bibitem[{\citenamefont{Achucarro and Urrestilla}(2000)}]{AchUrr}
\bibinfo{author}{\bibfnamefont{A.}~\bibnamefont{Achucarro}} \bibnamefont{and}
  \bibinfo{author}{\bibfnamefont{J.}~\bibnamefont{Urrestilla}},
  \bibinfo{journal}{Phys. Rev. Lett.} \textbf{\bibinfo{volume}{85}},
  \bibinfo{pages}{3091} (\bibinfo{year}{2000}), \eprint{hep-ph/0003145}.

\bibitem[{\citenamefont{Rhie and Bennett}(1991)}]{Rhie}
\bibinfo{author}{\bibfnamefont{S.~H.} \bibnamefont{Rhie}} \bibnamefont{and}
  \bibinfo{author}{\bibfnamefont{D.~P.} \bibnamefont{Bennett}},
  \bibinfo{journal}{Phys. Rev. Lett.} \textbf{\bibinfo{volume}{67}},
  \bibinfo{pages}{1173} (\bibinfo{year}{1991}).

\bibitem[{\citenamefont{Perivolaropoulos}(1992)}]{Perivolaropoulos}
\bibinfo{author}{\bibfnamefont{L.}~\bibnamefont{Perivolaropoulos}},
  \bibinfo{journal}{Nucl. Phys.} \textbf{\bibinfo{volume}{B375}},
  \bibinfo{pages}{665} (\bibinfo{year}{1992}).

\bibitem[{\citenamefont{Bennett and Rhie}(1990)}]{BRHIE}
\bibinfo{author}{\bibfnamefont{D.~P.} \bibnamefont{Bennett}} \bibnamefont{and}
  \bibinfo{author}{\bibfnamefont{S.~H.} \bibnamefont{Rhie}},
  \bibinfo{journal}{Phys. Rev. Lett.} \textbf{\bibinfo{volume}{65}},
  \bibinfo{pages}{1709} (\bibinfo{year}{1990}).

\bibitem[{\citenamefont{Yamaguchi}(2002)}]{YAMAGUCHI}
\bibinfo{author}{\bibfnamefont{M.}~\bibnamefont{Yamaguchi}},
  \bibinfo{journal}{Phys. Rev.} \textbf{\bibinfo{volume}{D65}},
  \bibinfo{pages}{063518} (\bibinfo{year}{2002}), \eprint{hep-ph/0107230}.

\bibitem[{\citenamefont{Bennett and Rhie}(1993)}]{cmb1}
\bibinfo{author}{\bibfnamefont{D.~P.} \bibnamefont{Bennett}} \bibnamefont{and}
  \bibinfo{author}{\bibfnamefont{S.~H.} \bibnamefont{Rhie}},
  \bibinfo{journal}{Astrophys. J.} \textbf{\bibinfo{volume}{406}},
  \bibinfo{pages}{L7} (\bibinfo{year}{1993}), \eprint{hep-ph/9207244}.

\bibitem[{\citenamefont{Durrer et~al.}(2002)\citenamefont{Durrer, Kunz, and
  Melchiorri}}]{cmb2}
\bibinfo{author}{\bibfnamefont{R.}~\bibnamefont{Durrer}},
  \bibinfo{author}{\bibfnamefont{M.}~\bibnamefont{Kunz}}, \bibnamefont{and}
  \bibinfo{author}{\bibfnamefont{A.}~\bibnamefont{Melchiorri}},
  \bibinfo{journal}{Phys. Rept.} \textbf{\bibinfo{volume}{364}},
  \bibinfo{pages}{1} (\bibinfo{year}{2002}), \eprint{astro-ph/0110348}.

\bibitem[{\citenamefont{Bevis et~al.}(2004)\citenamefont{Bevis, Hindmarsh, and
  Kunz}}]{cmb3}
\bibinfo{author}{\bibfnamefont{N.}~\bibnamefont{Bevis}},
  \bibinfo{author}{\bibfnamefont{M.}~\bibnamefont{Hindmarsh}},
  \bibnamefont{and} \bibinfo{author}{\bibfnamefont{M.}~\bibnamefont{Kunz}},
  \bibinfo{journal}{Phys. Rev.} \textbf{\bibinfo{volume}{D70}},
  \bibinfo{pages}{043508} (\bibinfo{year}{2004}), \eprint{astro-ph/0403029}.

\bibitem[{\citenamefont{Martins and Shellard}(1996)}]{MS1}
\bibinfo{author}{\bibfnamefont{C.~J. A.~P.} \bibnamefont{Martins}}
  \bibnamefont{and} \bibinfo{author}{\bibfnamefont{E.~P.~S.}
  \bibnamefont{Shellard}}, \bibinfo{journal}{Phys. Rev.}
  \textbf{\bibinfo{volume}{D54}}, \bibinfo{pages}{2535} (\bibinfo{year}{1996}),
  \eprint{hep-ph/9602271}.

\bibitem[{\citenamefont{Martins and Shellard}(2002)}]{MS2}
\bibinfo{author}{\bibfnamefont{C.~J. A.~P.} \bibnamefont{Martins}}
  \bibnamefont{and} \bibinfo{author}{\bibfnamefont{E.~P.~S.}
  \bibnamefont{Shellard}}, \bibinfo{journal}{Phys. Rev.}
  \textbf{\bibinfo{volume}{D65}}, \bibinfo{pages}{043514}
  (\bibinfo{year}{2002}), \eprint[http://arXiv.org/abs]{hep-ph/0003298}.

\bibitem[{\citenamefont{Preskill}(1979)}]{PRESKILL}
\bibinfo{author}{\bibfnamefont{J.}~\bibnamefont{Preskill}},
  \bibinfo{journal}{Phys. Rev. Lett.} \textbf{\bibinfo{volume}{43}},
  \bibinfo{pages}{1365} (\bibinfo{year}{1979}).

\bibitem[{\citenamefont{Kibble}(1985)}]{KIB}
\bibinfo{author}{\bibfnamefont{T.~W.~B.} \bibnamefont{Kibble}},
  \bibinfo{journal}{Nucl. Phys.} \textbf{\bibinfo{volume}{B252}},
  \bibinfo{pages}{227} (\bibinfo{year}{1985}).

\bibitem[{\citenamefont{Bennett}(1986)}]{BMOD}
\bibinfo{author}{\bibfnamefont{D.~P.} \bibnamefont{Bennett}},
  \bibinfo{journal}{Phys. Rev.} \textbf{\bibinfo{volume}{D33}},
  \bibinfo{pages}{872} (\bibinfo{year}{1986}).

\bibitem[{\citenamefont{Austin et~al.}(1993)\citenamefont{Austin, Copeland, and
  Kibble}}]{ACK}
\bibinfo{author}{\bibfnamefont{D.}~\bibnamefont{Austin}},
  \bibinfo{author}{\bibfnamefont{E.~J.} \bibnamefont{Copeland}},
  \bibnamefont{and} \bibinfo{author}{\bibfnamefont{T.~W.~B.}
  \bibnamefont{Kibble}}, \bibinfo{journal}{Phys. Rev.}
  \textbf{\bibinfo{volume}{D48}}, \bibinfo{pages}{5594} (\bibinfo{year}{1993}),
  \eprint{hep-ph/9307325}.

\bibitem[{\citenamefont{Moore et~al.}(2002)\citenamefont{Moore, Shellard, and
  Martins}}]{ABELIAN}
\bibinfo{author}{\bibfnamefont{J.~N.} \bibnamefont{Moore}},
  \bibinfo{author}{\bibfnamefont{E.~P.~S.} \bibnamefont{Shellard}},
  \bibnamefont{and} \bibinfo{author}{\bibfnamefont{C.~J. A.~P.}
  \bibnamefont{Martins}}, \bibinfo{journal}{Phys. Rev.}
  \textbf{\bibinfo{volume}{D65}}, \bibinfo{pages}{023503}
  (\bibinfo{year}{2002}), \eprint{hep-ph/0107171}.

\bibitem[{\citenamefont{Martins et~al.}(2004)\citenamefont{Martins, Moore, and
  Shellard}}]{MS3}
\bibinfo{author}{\bibfnamefont{C.~J. A.~P.} \bibnamefont{Martins}},
  \bibinfo{author}{\bibfnamefont{J.~N.} \bibnamefont{Moore}}, \bibnamefont{and}
  \bibinfo{author}{\bibfnamefont{E.~P.~S.} \bibnamefont{Shellard}},
  \bibinfo{journal}{Phys. Rev. Lett.} \textbf{\bibinfo{volume}{92}},
  \bibinfo{pages}{251601} (\bibinfo{year}{2004}), \eprint{hep-ph/0310255}.

\bibitem[{\citenamefont{Martins and Shellard}(2006)}]{MS4}
\bibinfo{author}{\bibfnamefont{C.~J. A.~P.} \bibnamefont{Martins}}
  \bibnamefont{and} \bibinfo{author}{\bibfnamefont{E.~P.~S.}
  \bibnamefont{Shellard}}, \bibinfo{journal}{Phys. Rev.}
  \textbf{\bibinfo{volume}{D73}}, \bibinfo{pages}{043515}
  (\bibinfo{year}{2006}), \eprint{astro-ph/0511792}.

\bibitem[{\citenamefont{Avelino et~al.}(2005)\citenamefont{Avelino, Martins,
  and Oliveira}}]{AWALL}
\bibinfo{author}{\bibfnamefont{P.~P.} \bibnamefont{Avelino}},
  \bibinfo{author}{\bibfnamefont{C.~J. A.~P.} \bibnamefont{Martins}},
  \bibnamefont{and} \bibinfo{author}{\bibfnamefont{J.~C. R.~E.}
  \bibnamefont{Oliveira}}, \bibinfo{journal}{Phys. Rev.}
  \textbf{\bibinfo{volume}{D72}}, \bibinfo{pages}{083506}
  (\bibinfo{year}{2005}), \eprint{hep-ph/0507272}.

\bibitem[{\citenamefont{Coleman}(1983)}]{LECTURESC}
\bibinfo{author}{\bibfnamefont{S.}~\bibnamefont{Coleman}}
  (\bibinfo{year}{1983}), \bibinfo{note}{in The Unity of Fundamental
  Interactions, A Zichichi et al. (eds.), Plenum Press New York}.

\bibitem[{\citenamefont{Preskill}(1985)}]{LECTURESP}
\bibinfo{author}{\bibfnamefont{J.}~\bibnamefont{Preskill}}
  (\bibinfo{year}{1985}), \bibinfo{note}{lectures presented at the 1985 Les
  Houches Summer School}.

\bibitem[{\citenamefont{Hindmarsh and Rajantie}(2000)}]{Hindmarsh:2000kd}
\bibinfo{author}{\bibfnamefont{M.}~\bibnamefont{Hindmarsh}} \bibnamefont{and}
  \bibinfo{author}{\bibfnamefont{A.}~\bibnamefont{Rajantie}},
  \bibinfo{journal}{Phys. Rev. Lett.} \textbf{\bibinfo{volume}{85}},
  \bibinfo{pages}{4660} (\bibinfo{year}{2000}), \eprint{cond-mat/0007361}.

\bibitem[{\citenamefont{Rajantie}(2003)}]{ARTTU}
\bibinfo{author}{\bibfnamefont{A.}~\bibnamefont{Rajantie}},
  \bibinfo{journal}{Phys. Rev.} \textbf{\bibinfo{volume}{D68}},
  \bibinfo{pages}{021301} (\bibinfo{year}{2003}), \eprint{hep-ph/0212130}.

\bibitem[{\citenamefont{Barriola and Vilenkin}(1989)}]{Barriola}
\bibinfo{author}{\bibfnamefont{M.}~\bibnamefont{Barriola}} \bibnamefont{and}
  \bibinfo{author}{\bibfnamefont{A.}~\bibnamefont{Vilenkin}},
  \bibinfo{journal}{Phys. Rev. Lett.} \textbf{\bibinfo{volume}{63}},
  \bibinfo{pages}{341} (\bibinfo{year}{1989}).

\bibitem[{\citenamefont{Einhorn et~al.}(1980)\citenamefont{Einhorn, Stein, and
  Toussaint}}]{EINHORN}
\bibinfo{author}{\bibfnamefont{M.~B.} \bibnamefont{Einhorn}},
  \bibinfo{author}{\bibfnamefont{D.~L.} \bibnamefont{Stein}}, \bibnamefont{and}
  \bibinfo{author}{\bibfnamefont{D.}~\bibnamefont{Toussaint}},
  \bibinfo{journal}{Phys. Rev.} \textbf{\bibinfo{volume}{D21}},
  \bibinfo{pages}{3295} (\bibinfo{year}{1980}).

\bibitem[{\citenamefont{Zeldovich and Khlopov}(1978)}]{KHLOPOV}
\bibinfo{author}{\bibfnamefont{Y.~B.} \bibnamefont{Zeldovich}}
  \bibnamefont{and} \bibinfo{author}{\bibfnamefont{M.~Y.}
  \bibnamefont{Khlopov}}, \bibinfo{journal}{Phys. Lett.}
  \textbf{\bibinfo{volume}{B79}}, \bibinfo{pages}{239} (\bibinfo{year}{1978}).

\bibitem[{\citenamefont{Achucarro et~al.}(2007)\citenamefont{Achucarro, Salmi,
  and Urrestilla}}]{Achucarro:2007sp}
\bibinfo{author}{\bibfnamefont{A.}~\bibnamefont{Achucarro}},
  \bibinfo{author}{\bibfnamefont{P.}~\bibnamefont{Salmi}}, \bibnamefont{and}
  \bibinfo{author}{\bibfnamefont{J.}~\bibnamefont{Urrestilla}},
  \bibinfo{journal}{Phys. Rev.} \textbf{\bibinfo{volume}{D75}},
  \bibinfo{pages}{121703} (\bibinfo{year}{2007}), \eprint{astro-ph/0512487}.

\bibitem[{\citenamefont{Achucarro and Vachaspati}(2000)}]{Achucarro:1999it}
\bibinfo{author}{\bibfnamefont{A.}~\bibnamefont{Achucarro}} \bibnamefont{and}
  \bibinfo{author}{\bibfnamefont{T.}~\bibnamefont{Vachaspati}},
  \bibinfo{journal}{Phys. Rept.} \textbf{\bibinfo{volume}{327}},
  \bibinfo{pages}{347} (\bibinfo{year}{2000}), \eprint{hep-ph/9904229}.

\bibitem[{\citenamefont{Achucarro et~al.}(1999)\citenamefont{Achucarro,
  Borrill, and Liddle}}]{Achucarro:1998ux}
\bibinfo{author}{\bibfnamefont{A.}~\bibnamefont{Achucarro}},
  \bibinfo{author}{\bibfnamefont{J.}~\bibnamefont{Borrill}}, \bibnamefont{and}
  \bibinfo{author}{\bibfnamefont{A.~R.} \bibnamefont{Liddle}},
  \bibinfo{journal}{Phys. Rev. Lett.} \textbf{\bibinfo{volume}{82}},
  \bibinfo{pages}{3742} (\bibinfo{year}{1999}), \eprint{hep-ph/9802306}.

\bibitem[{\citenamefont{Urrestilla et~al.}(2002)\citenamefont{Urrestilla,
  Achucarro, Borrill, and Liddle}}]{Urrestilla:2001dd}
\bibinfo{author}{\bibfnamefont{J.}~\bibnamefont{Urrestilla}},
  \bibinfo{author}{\bibfnamefont{A.}~\bibnamefont{Achucarro}},
  \bibinfo{author}{\bibfnamefont{J.}~\bibnamefont{Borrill}}, \bibnamefont{and}
  \bibinfo{author}{\bibfnamefont{A.~R.} \bibnamefont{Liddle}},
  \bibinfo{journal}{JHEP} \textbf{\bibinfo{volume}{08}}, \bibinfo{pages}{033}
  (\bibinfo{year}{2002}), \eprint{hep-ph/0106282}.

\end{thebibliography}
\end{document}